\documentstyle[12pt]{article}

\advance\textheight by \topskip
\newcommand{\Dir}{\kern -6.4pt\Big{/}}
\newcommand{\Dirin}{\kern -13.4pt\Big{/}\kern 6.4pt}
\newcommand{\DDir}{\kern -7.6pt\Big{/}}
\newcommand{\DGir}{\kern -6.0pt\Big{/}}

\begin{document}

\thispagestyle{empty}
\setcounter{page}{0}

\begin{flushright}
{\large DFTT 35/93}\\
{\rm June 1993\hspace*{.5 truecm}}\\
\end{flushright}

\vspace*{\fill}

\begin{center}
{\Large \bf Pion Compton Scattering in Perturbative QCD
\footnote{ Work supported in part by Ministero
dell' Universit\`a e della Ricerca Scientifica.}}\\[2cm]
{\large Ezio Maina and Roberto Torasso }\\[.3 cm]
{\it Dipartimento di Fisica Teorica, Universit\`a di Torino, Italy}\\
{\it and INFN, Sezione di Torino, Italy}\\[1cm]
\end{center}

\vspace*{\fill}

\begin{abstract}
{\normalsize
Pion Compton scattering is studied in perturbative QCD for real and space--like
initial photons. Different methods for the convolution of the hard amplitude
with
the pion wave--functions, which have in the past led to conflicting results,
are compared.
}
\end{abstract}

\vspace*{\fill}

\newpage
\subsection*{Introduction}
The analysis of exclusive processes in QCD has vastly
improved in the last few years. The r\^ole of Sudakov logarithms
in elastic scattering has been investigated \cite{BS}. It has been shown
\cite{FSZ} that,
like form factors and $\gamma\gamma$ reactions, large angle photoproduction
and Compton scattering are not affected in lowest order by Sudakov corrections
and that, as a consequence, the magnitude and the phase of the scattering
amplitudes for
these processes can be reliably predicted by perturbative QCD (pQCD) at large
momentum transfer. The applicability of pQCD to hadron form factor has been
extended to
lower energies using an improved version of the factorization formula
\cite{LiS}.
The sum--rule approach has been applied to Compton scattering \cite{COR1,COR2}.
\par
Nonetheless, the number of reactions for which theoretical predictions
exist is still rather small. The difficulties are best exemplified by the
existence of three published calculations for proton Compton scattering,
$\gamma p \rightarrow \gamma p$, in disagreement with each other
\cite{MF,KN,FZ}.\par
In this paper we consider pion Compton scattering for
real and space--like initial photons in pQCD. Pion Compton is the simplest
reaction
in which non trivial--phases are present and which is
experimentally accessible. Its calculation is conceptually similar
to the calculation of proton Compton scattering, but much simpler.
It is, therefore, an ideal laboratory for studying the different methods
\cite{MF,KN,FZ}
for the convolution of the hard amplitude.\par
Pion Compton has been considered before for real \cite{COR2,TAMAZ} and
time--like photons \cite{TAMAZ}, in this paper we extend
previous calculations to new kinematical domains and correct some errors that
are present in \cite{TAMAZ}.\par
There are different ways
of measuring the $\pi\gamma$ cross section. In the first instance a beam of
pions
is scattered off  of a nucleus.  In the forward direction the
one--photon--exchange
mechanism dominates \cite{bdef} and $\pi\gamma$ scattering can be studied.
Alternatively the interaction can be between an high--energy electron beam and
a nucleus,
with the extraction of quasi--real pions from nuclear matter.
\subsection*{Calculation}
Computing the amplitude for an exclusive process involves two steps.
First of all, an analytical expression for the large set of diagrams which is
often required needs to be derived. This is usually far from trivial
and in some cases can only be accomplished by some automatic procedure.
Second, the result has to be multiplied by the wave functions of all external
hadrons and then it has to be integrated over the momentum fractions of
all external quarks. Apart from simple cases like lepton--hadron scattering
and photon--photon annihilation, for particular values of the quark momenta,
one or more of the particles propagating inside the various diagrams can go on
their mass--shell.
Mathematically this means that one--dimensional integrals of the form:
\begin{equation}\label{basic}
\lim_{\epsilon\rightarrow 0}\int_0^1{f(x)\over x-a+i\epsilon}
\,dx=P\!\int_{0}^{1}{f(x)\over x-a}\,dx-i\pi f(a)
\end{equation}
where $0 < a < 1$, have to be computed, and that the result has to be
integrated
over all remaining variables.\par
Several methods have been proposed but unfortunately when applied to the
reaction $\gamma p \rightarrow \gamma p \,$
they have led to conflicting results.
Therefore we have preferred not to rely on only one method of integration.
This also allows us to compare the different procedures and to study their
relative merits.\par
We have considered a simpler version of the method used in \cite{MF},
which we will call the subtraction method, the technique suggested in
\cite{KN}, which we will call the KN method and the procedure
proposed in \cite{FZ}, which will be referred to as the $i\epsilon$ method.\par
In the first two cases some transformation is applied to the principal part
integral in the right--hand side of
(\ref{basic}) in order to avoid the large cancellations involved
which lead to large numerical instabilities. The imaginary part is then
computed
by hand, a procedure that is rather tedious and that, if not automated, can
be a source of errors which are almost impossible to detect.\par
The subtraction method is based on the following identity:
\begin{eqnarray}
P\! \int_{0}^{1}{f(x)\over x-a}\,dx & = &
\int_{0}^{1}{f(x)-f(a)\over x-a}\,
dx+f(a)\, P\! \int_{0}^{1}{1\over x-a}\,dx \nonumber \\
 & = &
 \int_{0}^{1}{f(x)-f(a)\over x-a}\,dx+f(a)\,\log{{1-a\over a}}
\end{eqnarray}
Both resulting integrals are numerically stable.\par
The KN method relies on a change of variable to simplify the
evaluation of the real part of eq. (\ref{basic}). One can write:
\begin{equation}
P\!\int_{0}^{1}{f(x)\over x-a}\,dx=\lim_{\epsilon\rightarrow 0}
(J_{1}+J_{2})
\end{equation}
where
\begin{equation}
J_{1}=\int_{0}^{a-\epsilon}{f(x)\over x-a}\,dx\qquad J_{2}=\int_{a
+\epsilon}^{1}{f(x)\over x-a}\,dx
\end{equation}
Now the following change of variables is performed on $J_1$:
\begin{equation}\label{eq:kn1}
y_{1}={x\over a}\Rightarrow J_{1}=\int_{0}^{1-\epsilon/a}{f(ay_{1})
\over y_{1}-1}\,dy_{1}
\end{equation}
and on $J_{2}$:
\begin{equation}\label{eq:kn2}
y_{2}={a(1-x)\over a^{2}+(1-2a)x}\Rightarrow J_{2}=\int_{0}^{1-
\epsilon/a}
{(1-a)f(g)\over d(1-y)}\,dy_{2}
\end{equation}
where
\begin{equation}
 g={a(1-ay)\over d}\qquad d=a+(1-2a)y
\end{equation}
The integration variables $y_{1}$, $y_{2}$ satisfy:
\begin{equation}
{dx\over dy_{1}}\big|_{y_{1}=1}=-{dx\over dy_{2}}\big|_{y_{2}=1}
\end{equation}
therefore the upper limits in (\ref{eq:kn1}, \ref{eq:kn2}) approach
$1$ uniformly. From (\ref{eq:kn1}, \ref{eq:kn2}) one can derive an expression
for the principal part of the integral in which the limit
$\epsilon\rightarrow0$ can be safely taken:

\begin{equation}
P\!\int_{0}^{1}{f(x)\over x-a}\,dx=
   \int_{0}^{1}\left\{{f(ay)\over y-1}+{(1-a)f(g)\over
 d(1-y)}\right\}\,dy
\end{equation}
The last integral is well behaved and can be directly integrated
numerically.\par
In our experience the subtraction method and the KN method always give the
same results and can be implemented with comparable effort.\par
The $i\epsilon$ method is the easiest one to implement automatically on
the large set of diagrams required to describe exclusive scatterings at tree
level. It has the additional advantage that the real and imaginary part of
eq. (\ref{basic}) are generated with the same algorithm and that, in principle,
the integrand once generated is not manipulated any further, hence no mistakes
are stirred in.
The method consists in evaluating numerically the real and imaginary
part of the integrals
\begin{equation}
\int_0^1{f(x)\over x-a+i\epsilon}\,dx
\end{equation}
for a number of small but finite values of $\epsilon$,
since for too small values of $\epsilon$ numerical instabilities take over,
and then to extrapolate to $\epsilon = 0$.
In \cite{FZ} Farrar and Zhang noted that for
$5.\times 10^{-4}<\epsilon < 1.\times 10^{-2}$ the
results seemed to depend very little on $\epsilon$ and therefore assumed that
the amplitudes obtained for $\epsilon = 5.\times 10^{-3}$ could be taken as a
good
approximation to the $\epsilon= 0$ results. This procedure was later criticized
in \cite{KN} since it failed to reproduce the correct answer in a number of
cases in which an analytical evaluation was feasible.\par
Our findings, for the much simpler set of diagrams we have been studying,
are in agreement with those in \cite{KN}. For
$\epsilon = 5.\times 10^{-3}$
the difference between the amplitudes evaluated with
the $i\epsilon$ method and with the two others can be as large as $4\div 5$\%
and $20\div 30$\%
for the real and imaginary part of the amplitude respectively.
For $\epsilon = 5.\times 10^{-4}$,
a value for which the integrals show no sign of numerical
problems, the difference can still be of several percent.
In most cases
a much better agreement, compatible with
the numerical errors, can be obtained
fitting a number of results in the range
$5.\times 10^{-4}<\epsilon < 5.\times 10^{-3}$ with a quadratic polynomial
in $\epsilon$ whose coefficients are determined by the least squares  method,
and assuming the value of the polynomial for $\epsilon=0$ as the true result.
Some caution is however needed. We have found instances, for perfectly
reasonable values of the scattering angle and of the photon virtuality,
in which  the result is disturbingly sensitive to the chosen range.
In some cases,
in order to reproduce the correct result, the fit has to be performed on
a different range, for instance  $1.\times 10^{-4}<\epsilon < 1.\times
10^{-3}$. The difference with the result obtained through a fit in the
range $5.\times 10^{-4}<\epsilon < 5.\times 10^{-3}$ can be as large as 10\%.
This obviously casts some doubts on the possibility of a completely
automatic implementation of the method.\par
As already noted \cite{Fprivate}, the limit
in which the mass $q^2=-Q^2<0$ of
the virtual photon vanishes is particularly tricky. For non--vanishing
photon mass a larger number of propagators can go on mass--shell and,
after partial fractioning, one discovers that different terms require
different ranges in $\epsilon$ to give an acceptable extrapolation
to $\epsilon=0$.
\subsection*{Results}
We define  the scattering kinematics for
$e(k,h_e) \pi (p) \rightarrow e(k^\prime,h_e) \pi (p^\prime)
\gamma(q^\prime,\lambda)$
as in \cite{FZ}.
The initial and final pion and the final photon have momenta:
\begin{eqnarray}
p & = & P(1,0,0,-1) \nonumber \\
p^\prime & = & Q^\prime(1,-\sin \theta,0,-\cos\theta) \\
q^\prime & = & Q^\prime(1,\sin \theta,0,\cos\theta) \nonumber
\end{eqnarray}
while momenta of the initial and final electrons are
\begin{eqnarray}
k & = & K(1,\cos\phi\sin\alpha,\sin\phi\sin\alpha,\cos\alpha) \\
k^\prime & = & K^\prime(1,\cos\phi\sin\alpha^\prime,
\sin\phi\sin\alpha^\prime,\cos\alpha^\prime)
\nonumber
\end{eqnarray}
The diagrams contributing to $e\pi$ scattering can be divided into two
sets. The first set includes all diagrams in which the final photon is emitted
from
the quark lines and will be referred to as the pion Compton set. In this case
for the
momentum of the intermediate photon we have
\begin{equation}
q  =  (\sqrt{P^2-Q^2},0,0,P)
\end{equation}
The second set includes all diagrams in which the final photon is emitted from
the
electron line.\par
Momentum conservation requires the following relationships among the parameters
\begin{eqnarray}
K\sin\alpha & = & K^\prime\sin\alpha^\prime \nonumber \\
P & = & K\cos\alpha - K^\prime\cos\alpha^\prime \\
P + K & = & 2 Q^\prime + K^\prime \nonumber \\
Q^2 & = & 4Q^\prime(P-Q^\prime) \nonumber
\end{eqnarray}
For a (virtual) photon propagating along the positive $z$-axis, the
polarization vectors
are defined in the standard way
\begin{equation}\label{eps}
\epsilon^\mu_{L,R}={1\over{\sqrt{2}}}(0,-1,\pm i,0) \hspace{.5in}
\epsilon^\mu_{\pm} =
       {1\over{\sqrt{2}}}(\epsilon^\mu_0 {\pm} \epsilon^\mu_3) =
       {1\over{\sqrt{2}}}(1,0,0,\pm 1)
\end{equation}
Writing the amplitude $A^\lambda_{h_e}$
for the pion Compton diagrams as:
\begin{equation}
A^\lambda_{h_e} = M_\nu g^{\mu\nu} V_{\mu\rho}\epsilon^{\rho\ast}/q^2
\end{equation}
where $M_\nu$ describes the emission of a virtual photon from the lepton line
and
$T_\mu= V_{\mu\rho}\epsilon^{\rho\ast}$ describes the scattering of this photon
off the pion,
and using the identity
\begin{equation}
-g^{\mu\nu}  =  \epsilon^\mu_R {\epsilon^\nu_R}^\ast + \epsilon^\mu_L
{\epsilon^\nu_L}^\ast
              - \epsilon^\mu_+ {\epsilon^\nu_-}^\ast - \epsilon^\mu_-
{\epsilon^\nu_+}^\ast
\end{equation}
one can write
\begin{equation}
-q^2 A = M_L T_L + M_R T_R - M_-T_+ - M_+ T_-
\end{equation}
Gauge invariance of the amplitude $T_\mu$ requires
\begin{eqnarray}
q\cdot T & = & \sqrt{P^2-Q^2}\, T^0 - P \, T^3 \,=\,\sqrt{P^2-Q^2}
\epsilon_0\cdot T + P \epsilon_3\cdot T \\
         & = & {1\over{\sqrt{2}}}\sqrt{P^2-Q^2} \,(T_+ + T_-) +
                      {1\over{\sqrt{2}}} P \, (T_+ - T_-) \,= \,0 \nonumber
\end{eqnarray}
It is therefore sufficient to give the amplitudes
$V_{ij} = V_{\mu\nu} \epsilon^{\mu}_i\epsilon^{\nu\ast}_j$
for incoming photons
with polarization $\epsilon_L, \, \epsilon_R$ and $\epsilon_+$.\par
All the computations have been made in the formalism of \cite{FN}.
Our results for the unintegrated amplitudes
coincide with those in \cite{TAMAZ}.
In all formulae we use the abbreviations
$c = \cos \theta/2$, $c_\alpha = \cos \alpha/2$,
$c_{\alpha^\prime} = \cos {\alpha^\prime}/2$,
$s = \sin \theta/2$, $s_\alpha = \sin \alpha/2$,
$s_{\alpha^\prime} = \sin {\alpha^\prime}/2$.\par
It is amusing to note that the
amplitude for $\pi\gamma_L\rightarrow\pi\gamma_R$ with real photons
has a very simple form
\begin{equation}
V_{LR} = C_0 (e_1 -e_2)^2 (x-y)\,s^2/c^2
\end{equation}
where $C_0$ is a constant, $x$ and $y$ are the momentum fractions
of the initial and final quark and
$e_1$ and $e_2$ are the charges of the two quark lines.
This is in agreement with a general theorem
on helicity amplitudes \cite{wang}.
\par
We present only the amplitudes for right-handed final photons, since all others
are related by parity. The helicity amplitudes $A^\lambda_{h_e}$
for the pion Compton contribution to $e\pi\rightarrow e\pi\gamma$,
with the definitions given in (\ref{eps}), are
\begin{eqnarray}\label{amp1}
A^R_- & = & c_{\alpha^\prime}c_{\alpha}V_{+R} +
s_{\alpha^\prime}s_{\alpha}V_{-R}
            - e^{i\phi}s_{\alpha^\prime}c_{\alpha}V_{RR}
            - e^{-i\phi}c_{\alpha^\prime}s_{\alpha}V_{LR} \\
A^R_+ & = & c_{\alpha^\prime}c_{\alpha}V_{+R} +
s_{\alpha^\prime}s_{\alpha}V_{-R}
            - e^{i\phi}c_{\alpha^\prime}s_{\alpha}V_{RR}
            - e^{-i\phi}s_{\alpha^\prime}c_{\alpha}V_{LR}
\end{eqnarray}
times the common factor
\begin{equation}
c_A = {{2 e \sqrt{2 K K^\prime}}\over{Q^2}}
\end{equation}
The amplitudes $B^\lambda_{h_e}$ for the diagrams in which the photon is
emitted from the electron line are
\begin{eqnarray}\label{amp2a}
B^R_- & = & K^\prime
        (e^{i\phi}s_{\alpha^\prime}s + c_{\alpha^\prime}c )
(e^{-i\phi}s_{\alpha^\prime}c - c_{\alpha^\prime}s )
                      /2k^\prime\cdot q^\prime \nonumber \\
      &   &  \hspace{.2in} \times \left( P  c_{\alpha^\prime}c_{\alpha} +
            Q^\prime (e^{i\phi}s_{\alpha^\prime}s + c_{\alpha^\prime}c )
                              (e^{-i\phi}s_{\alpha}s + c_{\alpha}c ) \right)
\nonumber \\
      &   & + (e^{-i\phi}s_{\alpha}c - c_{\alpha}s ) /(-2k\cdot q^\prime)
\times \left[ P c_{\alpha^\prime} (
                                K c_{\alpha} ( e^{i\phi}s_{\alpha}s +
c_{\alpha}c ) - Q^\prime c ) \right. \\
      &   & \left. \hspace{.2in} + Q^\prime  (e^{i\phi}s_{\alpha^\prime}s +
c_{\alpha^\prime}c ) \left(
                                K ( e^{i\phi}s_{\alpha}s + c_{\alpha}c )
                       ( e^{-i\phi}s_{\alpha}s + c_{\alpha}c ) - Q^\prime
                              \right) \right]
\vspace{.2in}
\nonumber \\
B^R_+ & = & K (e^{i\phi}s_{\alpha}s + c_{\alpha}c ) (e^{-i\phi}s_{\alpha}c -
c_{\alpha}s )
                      /(-2k \cdot q^\prime)  \nonumber \\
      &   &  \hspace{.2in} \times \left( P  c_{\alpha^\prime}c_{\alpha} +
            Q^\prime (e^{-i\phi}s_{\alpha^\prime}s + c_{\alpha^\prime}c )
                              (e^{i\phi}s_{\alpha}s + c_{\alpha}c ) \right)
\nonumber \\
      &   & + (e^{-i\phi}s_{\alpha^\prime}c - c_{\alpha^\prime}s )
/2k^\prime\cdot q^\prime
                              \times \left[ P c_{\alpha} \left(
                                K^\prime c_{\alpha^\prime}
                             ( e^{i\phi}s_{\alpha^\prime}s + c_{\alpha^\prime}c
) + Q^\prime c
                                    \right) \right. \\
      &   & \left. \hspace{.2in} + Q^\prime  (e^{i\phi}s_{\alpha}s +
c_{\alpha}c ) (
                                K^\prime \left( e^{i\phi}s_{\alpha^\prime}s +
c_{\alpha^\prime}c )
                                   ( e^{-i\phi}s_{\alpha^\prime}s +
c_{\alpha^\prime}c ) + Q^\prime
                           \right) \right]
\nonumber
\end{eqnarray}
times the common factor
\begin{equation}
c_B = {{2^4 e^3 \sqrt{2 K K^\prime}}\over{t_{p p^\prime}}}\,F_\pi(t_{p
p^\prime})
\end{equation}
where $t_{p p^\prime}= -2 p\cdot p^\prime$ and $F_\pi$ is the pion form factor.
The full amplitude  is simply given by
$A^\lambda_{h_e}+B^\lambda_{h_e}$. Combining our results and the data for the
pion
electromagnetic form factor a prediction for the high energy
reaction $e\pi\rightarrow e\pi\gamma$ can be obtained.\par
We have tried three different forms for the pion wave--function
denoted CZ \cite{CZ1}, P2 and P3 \cite{FHZ}:
\begin{eqnarray}\label{wf}
\phi_{CZ}(x) & = & 30f_\pi x(1-x)(2x-1)^2 \nonumber \\
\phi_{P2}(x) & = & 6f_\pi x(1-x)\left[-0.1821+5.91(2x-1)^{2}\right] \\
\phi_{P3}(x) & = & 6f_\pi x(1-x)\left[0.6016-4.659(2x-1)^{2}
+15.52(2x-1)^{4}\right] \nonumber
\end{eqnarray}
where $f_\pi = .133$ GeV. For the strong coupling constant we have adopted
the traditional value $\alpha_s= .3$.
The expressions (\ref{wf}) for the wave function have all
been derived with the sum rule method \cite{SUM,CZ2}.\par
The amplitudes for $\pi\gamma_R\rightarrow\pi\gamma_R\,(RR)$,
$\pi\gamma_L\rightarrow\pi\gamma_R\,(LR)$ and
$\pi\gamma_+\rightarrow\pi\gamma_R\,(+R)$ are given in fig.1 through 3
for different values of the photon virtuality $\eta = Q^2/4{Q^\prime}^2$.
All other amplitudes can be obtained through parity transformations.
The corresponding phases are given in fig.4 through 6.\par
The amplitudes show a complex dependence on the scattering angle and
from the photon virtuality.
The $+R$ amplitude grows as expected with $\eta$ and only for
$\eta > .5$ gives a significant contribution. It peaks in the forward
and backward direction and otherwise has a smooth behaviour with a
phase close to $180^\circ$.
The $LR$ amplitude  is large in the backward direction and small
for small scattering angles as a consequence of angular momentum conservation.
It has a complicated dependence on $\eta$, decreasing for a while for
increasing values of
$\eta$ and then reversing the trend.
The $RR$ amplitude is the only one
with a non trivial phase for a real incoming photon
and gives the largest contribution.  It monotonically increases from small
to large scattering angle for real photons. For virtual photons it presents a
minimum at
$\cos\theta \simeq -0.3$, which is quite deep for the CZ wave--function, and
becomes
much larger than the amplitude for real photons in the forward direction.
Fig.4 shows that the minimum corresponds to a zero of the real part of the
amplitude.
The $RR$ contribution decreases rather smoothly with increasing $\eta$. \par
Our results for the $LR$ amplitude and for the real part of the $RR$ amplitude
agree with the results of \cite{TAMAZ}, which neglects all imaginary
contributions.\par
The predictions do not depend too strongly on the choice of the wavefunction
within the
limited set which has been considered.
With the exception of the $RR$ amplitude for the CZ wave--function and large
virtualities, the amplitudes squared
change at most by a factor of two or three for different wave--functions,
in agreement with past experiences. The phases are also
only moderately dependent upon the wave--function.\par
\subsection*{Conclusions}
We have computed pion Compton scattering in perturbative QCD for real and
space--like
initial photons.\par
We have studied three different methods for the convolution of the
hard amplitude with the pion wave--functions. All procedures give the same
results
when properly applied. We find that a careful extrapolation to $\epsilon=0$
is needed in the $i\epsilon$ method. This may partially explain some
disagreement between calculations performed in the past.\par
\subsection*{Acknowledgements}
We wish to express our gratitude to G.R. Farrar and H. Zhang who independently
computed
the diagrams of the Compton set. We have checked that the two results are in
complete agreement.

\vspace*{\fill}

\newpage
\subsection*{Figure Captions}
\begin{description}

\item[fig.1.] $s^3{{ {\it d}\sigma}\over {{\it d}\cos\theta}}$
for
$\pi\gamma_R\rightarrow\pi\gamma_R$ in the $\pi\gamma$
center--of--mass
for the three different wave--functions (a:CZ, b:P2, c:P3)
and for different values of the photon virtuality $\eta = Q^2/4{Q^\prime}^2$.

\item[fig.2.]  Same as in fig.1 for
$\pi\gamma_L\rightarrow\pi\gamma_R$.

\item[fig.3.]  Same as in fig.1 for
$\pi\gamma_+\rightarrow\pi\gamma_R$.

\item[fig.4.] Phase in degrees for
$\pi\gamma_R\rightarrow\pi\gamma_R$ as a function of the
scattering angle in the $\pi\gamma$
center--of--mass
for the three different wave--functions (a:CZ, b:P2, c:P3)
and for different values of the photon virtuality.

\item[fig.5.]  Same as in fig.4 for
$\pi\gamma_L\rightarrow\pi\gamma_R$.

\item[fig.6.]  Same as in fig.4 for
$\pi\gamma_+\rightarrow\pi\gamma_R$.

\end{description}

\end{document}